\documentclass[12pt,preprint]{aastex}
\usepackage{emulateapj5}

\def\ref{\par \noindent \hang}

\shorttitle{Subarcsec emission in Seyfert galaxies}
\shortauthors{Prieto et al.}

\begin{document}

\title{Subarcsec emission in Seyfert galaxies: the nuclear component in the 
$L$- and $M$-bands}

\author{M.A. Prieto}
\affil{European Southern Observatory, Karl-Schwarzschild-Str. 2, D--85748 
Garching, Germany}
\email{aprieto@eso.org}

\and

\author{J. Reunanen and J.K. Kotilainen}
\affil{Tuorla Observatory, University of Turku, V\"ais\"al\"antie 20, 
FIN--21500 Piikki\"o, Finland}
\email{reunanen@astro.utu.fi; jarkot@astro.utu.fi}

\begin{abstract}

We present deep $L$- (3.8 $\mu$m) and $M$- (4.7 $\mu$m) band imaging 
with {\it ISAAC} on the {\it ESO VLT} with unprecedented spatial 
resolution of the nearby Seyfert 2 galaxies NGC 7496 and NGC 7582 and 
the Seyfert 1 galaxy NGC 7213, all at comparable distance. 
The unresolved nuclear component dominates the emission within the central 
90 pc region, while the host galaxy accounts for up to 50 \% of 
the integrated emission at both wavelengths within the detected sizes of
of $\sim$1 kpc in the $L$-band and $\sim$0.5 kpc in the $M$-band. 
The overall morphology of the extended component follows the general 
isophote pattern defined by the near-infrared continuum of the 
galaxies. However, the central 300 pc regions show much more 
ordered elliptical isophotes than in the near-infrared. 
In particular, emission in the $L$- and $M$-bands shows well defined 
central point sources in the two Seyfert 2s. The brightness of the 
Seyfert nuclei in the $LM$-bands indicates that the Seyfert 2 
NGC 7582 harbors as powerful nucleus as that in the Seyfert 1 NGC 7213, 
while the Seyfert 2 NGC 7496 is intrinsically weaker. In NGC 7582, we 
have detected the circumnuclear star forming ring in the $L$-band. 

\end{abstract}


\section{Introduction}

The unified model for Seyfert galaxies postulates that Seyfert type 1 
(S1) and type 2 (S2) nuclei are intrinsically similar, the 
observed differences between them being due to obscuration and viewing 
angle to the AGN. The nuclear continuum and the broad line region (BLR) 
of S1s are obscured from direct view in S2s by an optically and 
geometrically thick torus of gas and dust. This torus absorbs a 
significant fraction of the nuclear high energy emission, and reradiates 
it into the infrared (IR; e.g. Antonucci 1993).

A promising method to test the validity of the unified model is to 
perform high spatial resolution imaging in the $LM$-bands, 
3 -- 5 $\mu$m. Recent near-IR (NIR; $JHK$-bands, 1 -- 2.5 $\mu$m) 
continuum imaging studies of Seyfert nuclei (e.g. 
Kotilainen \& Ward 1994; Alonso-Herrero, Ward \& Kotilainen 1996) have 
shown that the stellar contribution decreases with increasing 
wavelength, dominating in the $JH$-bands but accounting only for 
$\sim$50 \% of the emission in the $K$-band, particularly in S2s. 
If this trend persists towards longer wavelengths, the  3 -- 5 $\mu$m range
should give 
us a much cleaner view of the nuclear region in Seyferts. Furthermore, 
with reduced extinction effects, that spectral region may start to
 reveal the presence, 
and even the structure, of the putative obscuring dusty torus or 
disk surrounding the Seyfert nuclei. The internal regions of the torus 
may reach high temperatures (T $\geq$ 1000 K) and thus radiate at 
NIR wavelengths in S1s whereas this emission may remain optically thick 
in S2s. However, dust emitting in the 3 -- 5 $\mu$m range  should arise from 
regions further away from the nucleus, being thus cooler 
(T = 600 - 800 K) and equally traceable in both Seyfert types. 
By modeling the $LM$-band luminosity profiles, it should be possible 
to assess the contributions from stellar and dust emission, which should 
be similar in both Seyfert types if the unified model is valid.

Previous studies in the 3 -- 5 $\mu$m range are scarce. Pioneering work by 
Kotilainen et al. (1992), Zhou, Wynn-Williams \& Sanders (1993) 
and Alonso-Herrero et al. (1996; 2001) lack sufficient spatial 
resolution, sensitivity and field-of-view to detect the host galaxies 
and therefore do not allow for an accurate separation of the central 
nuclear component. Also, the available imaging only allows photometry 
with large apertures ($\geq$3{\arcsec}). Recently, 
diffraction-limited imaging in the $LM$-bands using adaptive optics (AO) 
of the nearby prototype S2 galaxy NGC 1068 revealed a wealth of information
(Marco \& Alloin 2000). These observations reveal in NGC 1068 a core 
with radius $<$ 8 pc, a disk structure of $\sim$80 pc size and 
extended emission perpendicular to the disk of $\sim$100 pc size.

This paper presents our first results 
from an  {\it ISAAC} program on the {\it ESO VLT} devoted to  subarcsec resolution images in the $LM$-bands of nearby S1 and S2 galaxies.
The results focuses on the  two S2s, NGC 7496 at z = 0.00550 and NGC 7582 at 
z = 0.00525, and one S1, NGC 7213 at z = 0.00597. The excellent 
image quality provided by {\it VLT} 
lead to unprecedentedly high spatial resolution images in 
the $LM$-bands, 0\farcs4 and 0\farcs6 FWHM, respectively, corresponding 
to physical distances between 70 and 90 pc.

\section{Observations}

Broad-band images of the Seyferts in the $L$- (3.8 $\mu$m) and $M$- 
(4.7 $\mu$m) bands were obtained with {\it ISAAC} on the {\it ESO VLT} 
UT {\it Antu} in service mode during 2000 September - October. 
The observations were taken in chopping mode. Each final image is the 
result of stacking 46 ($L$) or 104 ($M$) individual chopped frames. 
In addition to chopping, consecutive frames were taken after nodding 
the telescope to typically 20{\arcsec} distance on the sky. 
The total integration time for all galaxies was 25 minutes in the 
$L$-band and 45 minutes in the $M$-band. 
The field-of-view of the images is 73{\arcsec}, thus, these are the 
first large-format array $LM$-band images of any Seyfert galaxies. 
Previously published images in these bands cover smaller field-of-views by 
a factor of $\geq$ 3.  The instrumental PSF shape was determined from 
images of standard stars taken close in time to the galaxy images. 
Occasional image elongations were seen in the core of 
the PSF caused by field instabilities produced by the secondary mirror 
during chopping. These image elongations are generally $<$ 0\farcs5 FWHM 
in both bands. Additional $JHK$-band images were taken with {\it SOFI} 
on the {\it ESO NTT}. These images have a typical seeing of 
1{\arcsec} FWHM and integration times of 75 seconds in $JH$- and 
100 seconds in $K$-band.

\section{Results}

Fig. 1 shows the $LM$-band images of the three galaxies. All exhibit 
extended emission, traceable up to $\sim$3 - 7{\arcsec} from the nucleus. 
In the $L$-band, the extent of this emission ranges from 500 pc in 
NGC 7496 to 1.2 kpc in NGC 7213, whereas in the $M$-band, it is more 
compact and extends from 150 pc in NGC 7496 to 500 pc in NGC 7582. 

The S1 galaxy NGC 7213 is a face-on SA0, with identical 
morphology from the optical to the $M$-band. At 8.4-GHz, 
it is unresolved, the core region 
being less than 70 pc FWHM (Thean et al. 2000).
In contrast, the optical 
images of the S2s show a distorted nuclear region caused by dust 
extinction and the presence of bars. The SBcb  NGC 7582 is crossed by 
a dust lane along its major axis (Regan \& Mulchaey 1999). H$\alpha$ 
imaging reveals a continuous distribution of HII regions along its 
central disk or bar (Hameed \& Devereux 1999). The optical and NIR 
emission of the SBbc NGC 7496 is dominated by a long bar along 
its major axis, and its central 20{\arcsec} emission shows a 
peanut-shaped bulge (Mulchaey, Regan \& Kundu 1997). The $LM$-band images 
of these two S2s (Fig. 1) show, however, rather ordered elliptical 
isophotes, particularly in NGC 7582. In NGC 7496, 
the peanut-shaped morphology disappears, the $L$-band image revealing 
instead symmetric extended emisssion at opposite sides of the 
nucleus, presumably from the central bar.
At 8.4-GHz (Thean et al.), NGC 7582 is resolved in a peak and diffuse
emission extending over $\sim$1 kpc size whereas NGC 7496 is
unresolved down to a region of 43 pc FWHM.

In the $LM$-bands the innermost 90 pc emission of all three 
galaxies  is however  dominated by a central point source. To disentangle the 
nuclear component from the host galaxy, the azimuthally averaged 
radial luminosity profiles from $J$- to $M$-band were deconvolved into 
an unresolved point source component (AGN) and a bulge component 
(host galaxy). In the $LM$-band images, the point source component 
was approximated by the Moffat fit derived from the star observed close 
in time, while the bulge component was approximated by a 
de Vaucouleurs law. For NGC 7582, the extended emission was 
also alternatively fit with an exponential disk. In the $JHK$-band 
images, the unresolved component was directly approximated from the 
profiles of field stars in the galaxy frames. As the extension of 
the galaxies in these bands is larger than in the $LM$-bands, an 
additional disk component was included to better model the extended emission.

The resulting fits for NGC 7213 are shown as an example in Fig. 2., while 
the magnitudes and the fractional contributions of the unresolved and 
the extended component, and the modeled and observed luminosities in 
the $JHKLM$ bands within different apertures for all Seyferts are given 
in Table 1. The contribution from the unresolved component increases 
with decreasing aperture and increasing wavelength. In the $J$-band, 
the emission is dominated by the host galaxy in the S1 NGC 7213 and in the 
S2 NGC 7496, and accounts for $\sim$50 \% in the S2 NGC 7582. 
In the $K$-band, the host galaxy contribution is already reduced 
to $\sim$50 \% level, while in the $LM$-bands, the emission is dominated 
by the unresolved component in all three Seyferts. 

The spectral energy distributions (SED) of the nuclear and stellar 
components from $J$- to $M$-band within a 1{\arcsec} radius 
aperture ($\sim$150 pc) are shown in Fig. 3., along with similar data on 
NGC 1068 in a 0\farcs6 aperture (Marco \& Alloin 2000). The $K$- to 
$M$-band SED of the unresolved nuclear component is in all galaxies well 
fit by a grey black-body 
with temperature T = 500 - 800 K. This hot dust resides in a region 
$<$ 70-90 pc from the nucleus. Note, however, that the $JH$-band SED of 
the unresolved component clearly departs from that followed at 
longer wavelengts. The $J$-$H$ and $H$-$K$ colors (cf. magnitudes in 
Table 1) indicate much hotter dust. For all galaxies, the best fit to 
the $JH$-band emission indicates a dust temperature in the range 
T = 800 - 1600 K, close to the graphite sublimation temperature. 
However, the temperature of this dust component may be much cooler if 
an additional contribution by bremsstrahlung from the cooling of 
nuclear photoionized gas at T = $10^4 K$ is considered (see e.g. 
Contini \& Viegas 2001).

Overall, the $J$-$H$, $H$-$K$ and $K$-$L$ colors of the unresolved 
component are much redder than those found in quasars 
(Hyland \& Allen 1982). This difference is probably due to the still 
dominant contribution from the host galaxy, particularly in the 
$JH$-bands, in the central 150 pc region (Table 1), which renders 
the estimate of the unresolved component subject to  
large uncertainties. The $K$-$L$ colors are also redder than those 
derived for the central region of NGC 1068 ($K$-$L$ = 1.8$\pm$0.2; 
Marco \& Alloin 2000), although note that the contribution from the 
host galaxy in the $K$-band was not removed in this galaxy. This situation 
is reversed when comparing the $L$-$M$ colors: whereas in our three 
Seyferts $L$-$M$ $\sim$0.7, in NGC 1068 $L$-$M$ = 1.6$\pm$0.4, 
indicating that the emission peak of the unresolved component is toward 
even longer wavelengths in NGC 1068.

On the other hand, the SED of the host galaxy shows an opposite trend 
from that of the unresolved component, namely a smoothly 
decreasing contribution from $J$- to $M$-band. The stellar 
contribution appears to be unimportant in the central 80 - 100 pc region 
in the $LM$-bands, whereas it dominates in the $JHK$-bands, and also 
at larger distances in all bands. The $J$-$H$ and $H$-$K$ colors 
(Table 1) are close to those of giants and supergiants (Koorneef 1983) and 
of normal spirals (Willner et al. 1984). The $K$-$L$ colors are much 
redder than in normal spirals but this is expected as the 
unresolved component dominates the nuclear $L$-band light.

Finally, at slightly larger scales (radius $>$ 100 pc), the $L$-band 
emission from the host galaxy of NGC 7582 deserves particular attention. 
The subtraction of a bulge model from the $L$-band image reveals 
residual clumps of emission between 150 and 300 pc radius along the 
major axis of the galaxy (Fig. 4). These clumps are located along 
the disk/bar of HII regions seen also in H$\alpha$ 
(Hameed \& Devereux 1999). These clumps probably represent individual 
dust cocoons heated by the underlying ring of powerful star forming 
regions to temperatures high enough to emit in the $L$-band and 
beyond. The $L$-$M$ colors of the clumps indicate relatively hot dust, 
T $>$ 800 K, but these estimates are rather uncertain as the colors 
are averaged over the entire galaxy between 150 and 300 pc radius.

\section{Conclusions}

Deep, high spatial resolution $L$- and $M$-band imaging of three 
nearby Seyferts (two S2s and one S1) has allowed for the first time at 
these wavelengths to separate the AGN from the host galaxy 
emission. Regardless of the Seyfert type, in all three galaxies 
the unresolved nuclear component dominates the $LM$-band emission within 
$<$ 90 pc diameter central region. The contribution of this component in 
the $JHK$-bands is, on the other hand, $<$ 50 \%. Together with literature 
AO data for NGC 1068, the  $K$- to $M$-band SEDs of the unresolved 
component are well accounted for by dust emission with temperature 
T = 500 - 800 K in all cases.

The expected signature for central obscuring material   - the putative
disk or torus -- is not seen in these galaxies. If typical sizes
of the obscuring region are in the 80 pc range (cf.  NGC 1068) the
resolution of our data is just at the limit to resolve it. Future AO
observations will provide a definitive answer.  Interesting, recent
10 $\mu$m imaging of the S2 NGC 7582 reveals a nuclear disk structure
just at the right position, about perpendicular to the radio axis of
the galaxy (Acosta-Pulido et al. 2002).  We note that contarily to the
relative high polarization, P$\sim$16\%, measured in NGC 1068,
and  interpreted as due to scatter nuclear light from
its hidden S1 nucleus (Antonucci, 1993), the reported values 
for these galaxies are very
low, $\sim <$1\% (Brindle et al. 1990), and probably not intrinsic to
the sources.

The contribution of the host galaxy in the $LM$-bands is negligible 
within the central 90 pc region, but increases with radius and accounts 
for $>$ 50 \% of the integrated $LM$-band emission. The total extent of 
the host galaxy emission in the $L$-band is $\sim$1 kpc in all 
three Seyferts, whereas the $M$-band emission is more compact, with 
average size $\sim$500 pc.

The overall morphology of the galaxies at these wavelengths is 
rather relaxed.  Particularly, the emission in the $L$-band contrasts 
with the distorted optical and NIR emission of the two S2s. The lack of 
a prominent nucleus in the optical images contrasts with the bright 
nucleus seen in the $LM$-band images of the S2s. In particular, the 
nucleus of NGC 7582 is as bright as that of the S1 NGC 7213 in the 
$LM$-bands. Note that NGC 7582 is heavily absorbed, with 
N(H) = $1.67\times 10^{23}$ cm$^{-2}$ (Turner \& Pounds 1989). 
Correcting for this absorption, its hard X-ray (2-10 keV) luminosity 
is $4\times 10^{42}$ erg s$^{-1}$, of the same order as that in 
NGC 7213 (Nandra \& Pounds 1994). Thus, the brightness of its nucleus in 
the $LM$-bands indicates that NGC 7582 harbors a nucleus as powerful as 
that in NGC 7213.  
In the radio domain, the core flux  of  NGC 7213  at  8.4-GHz is
$\sim$ 3.5 larger than that measured in 
NGC 7582 in this case over a region of 10 arcsec  (cf. section 2);  within the
framework of the Seyfert Unifying Schemes the 
difference could be interpreted as due to emission from the jet
pointing close to the line of sight in the case of NGC 7213.

The S2 NGC 7496, although at the same distance as the other Seyferts, 
appears to be intrinsically weaker. It is also heavily absorbed, with 
N(H) $\sim 5\times 10^{22}$ cm$^{-2}$ (Kruper, Canizares \& Urry 1990), 
but the derived absorption-corrected hard X-ray (2-10 keV) luminosity 
is only $\sim 1\times 10^{42}$ erg s$^{-1}$, much lower than in the 
other cases. Its 8.4-GHz core emission   is also 
$\sim$ 14 times lower than in NGC 7582.
The  $LM$-band luminosities of the 
unresolved component  are also $\sim$3 magnitudes fainter than in 
the other two Seyferts, altogether confirming  NGC 7496 as the weakest
of the three AGN.

Finally, for NGC 7582 we report to our knowledge the first detection of 
a star forming ring around a Seyfert nucleus in the $L$-band. 
This detection highlights the potential of high spatial resolution 
studies in the 3 -- 5 $\mu$m range to reveal important  circunuclear 
emission in AGN.

\acknowledgments

Based on observations collected at the European Southern Observatory, 
Chile. We acknowledge the staff at Paranal Observatory, C. Lidman, 
J.G. Cuby and O. Marco, who carried out this {\it VLT ISAAC} program 
in service mode. We kindly thank R. Falomo for providing us with 
the luminosity profile fitting routines.

\begin{table}
\scriptsize
\caption{Fitting results at different wavelengths in three apertures.}
\begin{tabular}{lllllllllll}
\tableline\tableline
& & \multicolumn{3}{c}{NGC 7213} & \multicolumn{3}{c}{NGC 7496} & \multicolumn{3}{c}{NGC 7582}\\
$\lambda$ & Comp & 1\farcs0 & 2\farcs0 & 4\farcs0 & 1\farcs0 & 2\farcs0 & 4\farcs0 & 1\farcs0 & 2\farcs0 & 4\farcs0\\
\tableline
  J& nucl &.../... &14.88/~8        &14.62/~4             &.../... &15.48/21     &15.30/12   &.../...  &12.56/61        &12.98/39     \\
   & host  &.../... &12.30/92        &11.23/96                &.../... &14.09/79     &13.18/88   &.../...  &13.59/39        &12.51/61     \\
   & total&... ... &12.21 12.19     &11.19 11.17             &... ... &13.82 13.81  &13.04 13.04&... ...  &12.56 12.05     &11.97 11.29  \\
  H& nucl &.../... &13.29/17        &13.11/~9          &.../... &14.46/26     &14.27/16   &.../...  &11.34/80        &11.24/62     \\
   & host  &.../... &11.50/83        &10.47/91                &.../... &13.37/74     &12.47/84   &.../...  &12.84/20        &11.76/38     \\
   & total&... ... &11.31 11.29     &10.38 10.35             &... ... &13.03 13.01  &12.28 12.28&... ...  &11.10 10.74     &10.72 10.12  \\
 K& nucl &.../... &11.62/39        &11.49/22                &.../... &13.27/44     &13.14/38   &.../...  & ~9.98/90 & ~9.87/79 \\
   & host  &.../... &11.16/61        &10.16/78                &.../... &12.96/56     &12.14/72   &.../...  &12.32/10    &11.26/21     \\
&total&... ...&10.61 10.60& 9.88 9.86&... ...&12.38 12.37&11.78 11.78&... ...& 9.87 9.66& 9.61 9.19\\
  L& nucl & ~8.44/91  & ~8.13/83     & ~8.05/69  &11.47/89     &11.27/79     &11.26/61 &      ~8.30/89   & ~8.08/77  & ~8.07/63  \\
   & host  &10.95/~9     & ~9.82/17     & ~8.91/31  &11.77/11     &12.67/21     &11.74/39 &      10.52/11         & ~9.42/23  & ~8.63/37  \\
   & total& 8.34 8.36  & 7.92 7.95    & 7.64 7.67 &11.34 11.38    &11.00 10.94   &10.72/10.62& 8.17 8.17 & 7.81 7.82  & 7.63 7.57  \\
  M& nucl & ~7.71/93     & ~7.50/87     & ~7.45/75 &.../... & .../... & .../... & ~7.63/95   & ~7.42/89  & ~7.37/77  \\
   & host&10.60/~7     & ~9.54/13     & ~8.65/25 &.../... & .../... & .../...  &10.86/~5      & ~9.65/11  & ~8.68/23  \\
   & total& 8.34 8.36    & 7.92 7.95    & 7.64 7.67 &... 11.09        &... 10.59        &... 10.53 & 7.58 7.59 & 7.29 7.30  & 7.09 7.10  \\
\tableline
\end{tabular}
\end{table}

\figcaption[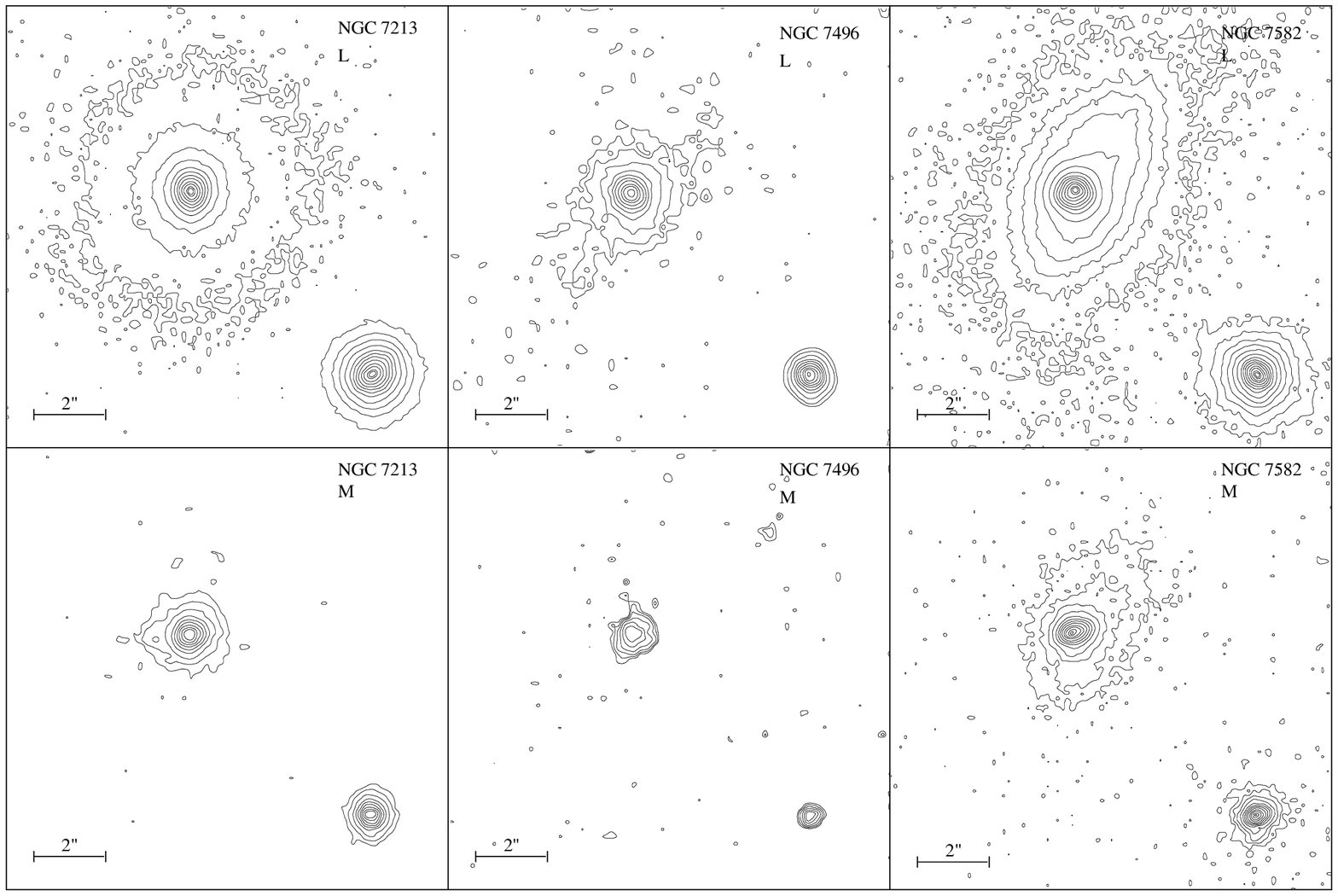]{Contour plots of the $L$- and $M$-band images of 
the three Seyferts. North is up and east to the left. The 2{\arcsec} 
scale bar, and contour plots of the standard stars observed for 
PSF determination are shown in the bottom left and bottom right corner 
of each panel, respectively.\label{fig1}}

\figcaption[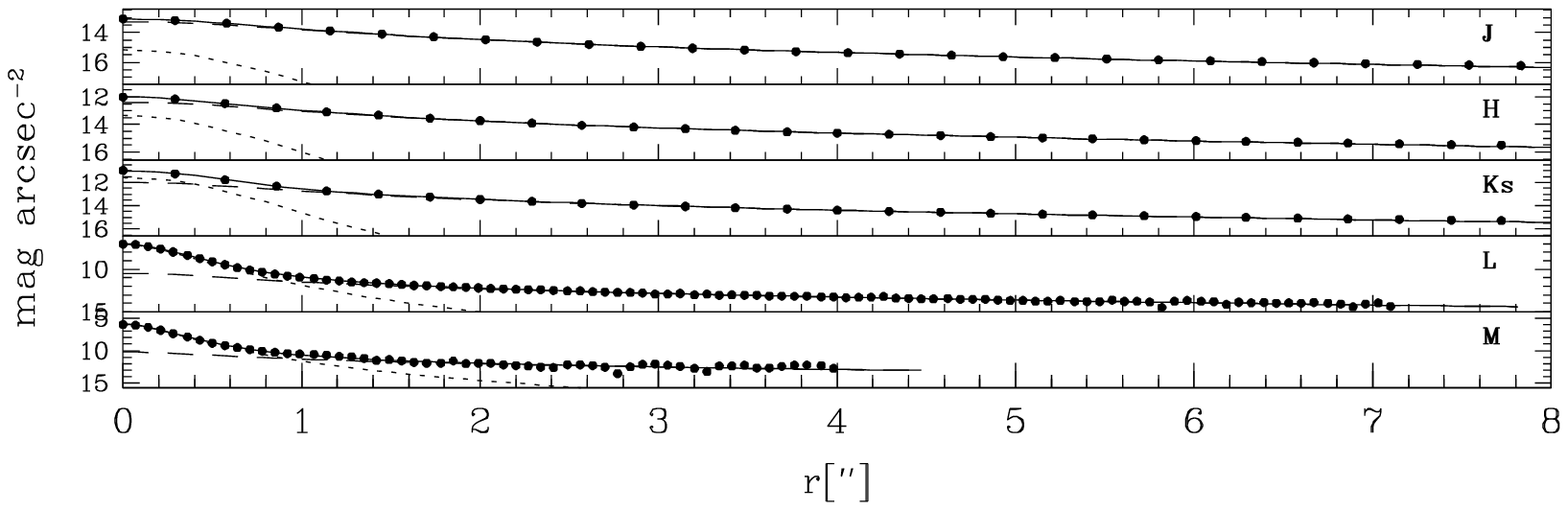]{Luminosity profile fits in the $J$- to $M$-bands of 
NGC 7213. Each panel shows the azimuthally averaged observed 
luminosity profile (filled circles), the contributions from the 
point source (short-dashed line) and bulge (long-dashed line) components, 
and the fit to the observed profile from the convolution of these 
components (solid line).\label{fig2}}

\figcaption[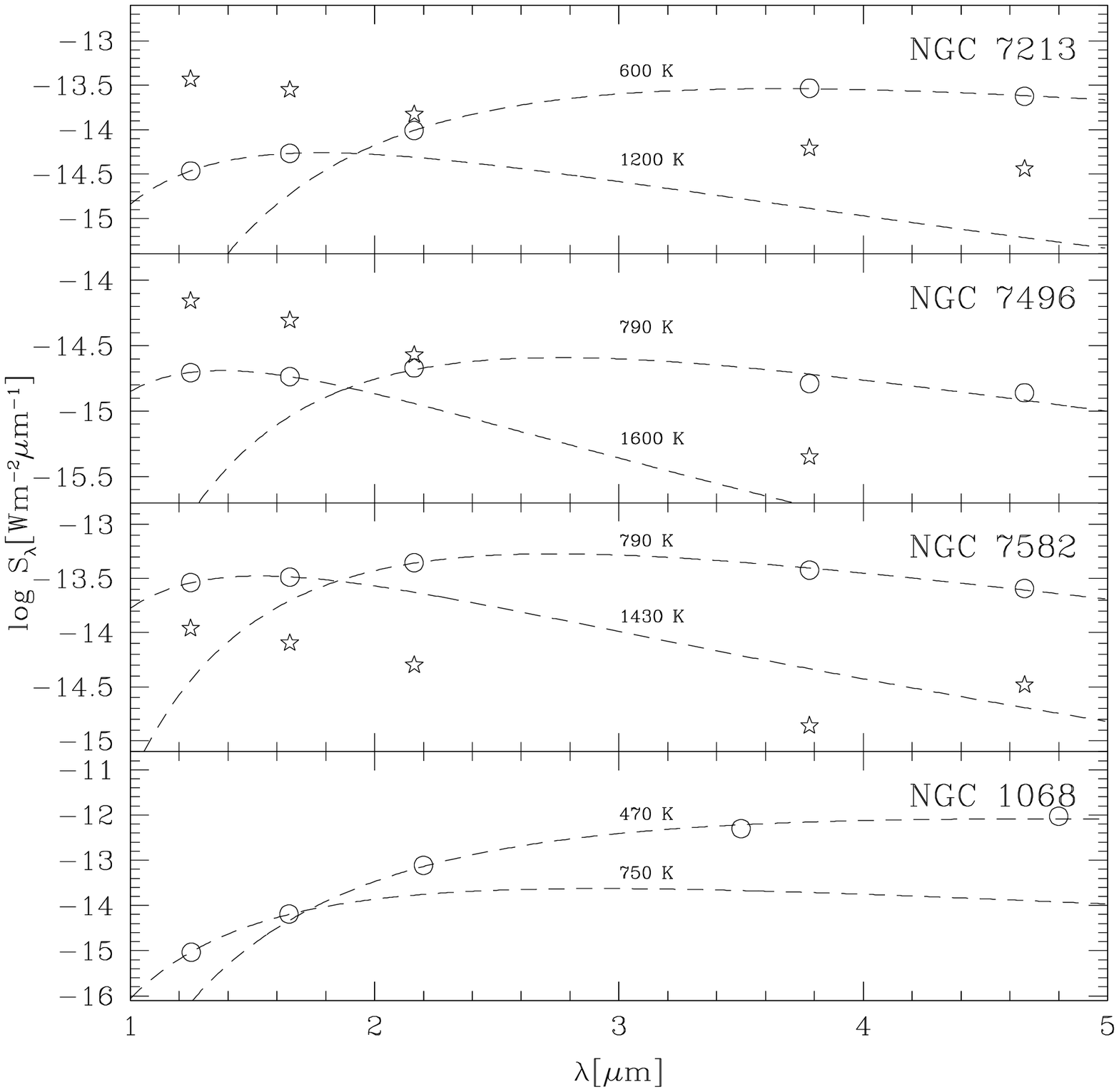]{The SED of the Seyferts from $J$- to $M$-band within 
a 1{\arcsec} radius aperture ($\sim$150 pc average size). 
The unresolved nuclear component and the host galaxy component are plotted 
as circles and stars, respectively. In the bottom panel are shown 
for comparison the results for NGC 1068 within a 0\farcs6 diameter 
radius aperture ($\sim$33 pc) from Marco \& Alloin (2000). For NGC 1068, 
all emission is assumed to be nuclear, as no deconvolution was applied 
in this case.\label{fig3}}

\figcaption[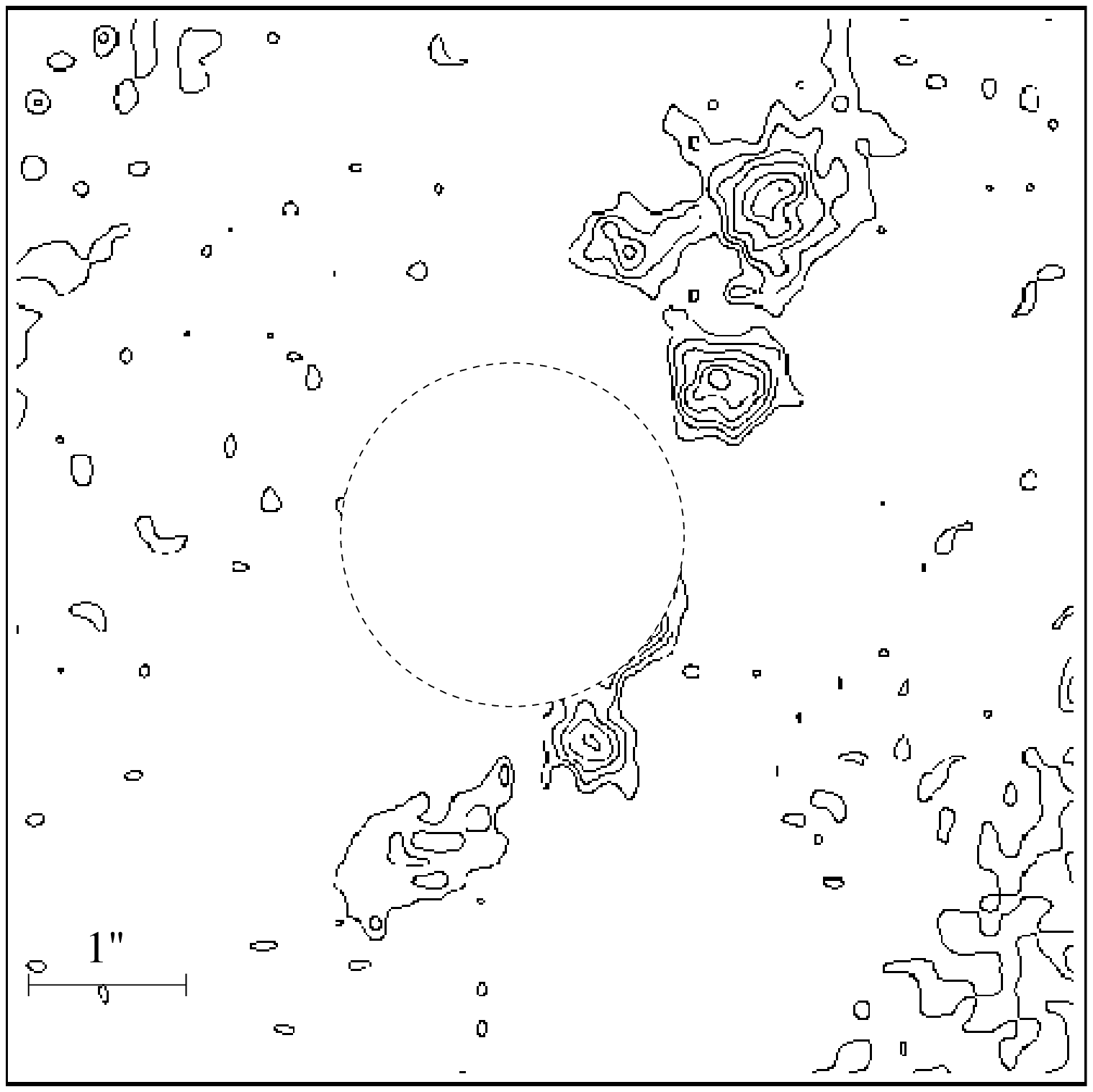]{The $L$-band image of NGC 7582 after removal of the 
host galaxy with a bulge model. Outside the central region, which is 
masked for the sake of clarity, the residual emission shows a ring 
of circunuclear clumps interpreted as powerful star forming 
regions.\label{fig4}}

\end{document}